ABSTRACT: With rising concerns about climate change, the issue of green growth have been growing in importance. The aim of this study is to establish a measurement method for green growth, and to identify the best performing countries in this field. The Future, Outside, and Inside (FOI) development model was used to measure the performance of the 38 OECD countries. Based on their 2019-20 scores, the countries that are top performers in green growth are the members of the so called Welfare-participatory cluster (Austria, Denmark, Finland, Germany, Ireland, Israel, New Zealand, Norway, Sweden), and two outliers (Iceland, and Luxembourg).

## 1 INTRODUCTION

As the topic of human-induced climate change came more and more to the forefront of public debate, the area that is often summarized as green growth, energy transition or (energy and/or growth) decoupling became one of the main research problems in economics. A ProQuest database search for content published in 2021 provides 589,822 hits for green growth (14% of the all-time value), 372,172 ones for energy transition (15%), and 19,408 ones for decoupling (10,5%). These abstractions refer to very similar concepts. Green growth is defined as "fostering economic growth and development, while ensuring that natural assets continue to provide the resources and environmental services on which our well-being relies" (OECD 2011, p. 9); energy transition refers to the transition from fossil fuels to renewable energy (Quitzow et al. 2019); while according to the OECD "decoupling occurs when the growth rate of an environmental pressure is less than that of its economic driving force (e.g. GDP)" (OECD 2002, p. 4). The big question is, whether green growth is happening, if the transition is happening quickly enough, and what are the policies that can facilitate the process.

This paper focuses on the OECD countries, uses the FOI model (Bartha & Gubik 2014) to evaluate their performance in the three main pillars of development (Future, Outside, and Inside potential), and investigates how green growth fits into the development model of the OECD countries. This study aims to answer two main questions: Is there a green growth pattern emerging among the OECD countries; and Which OECD countries perform the best in green growth? The contribution of this study is the following:

1. By measuring the Future, Outside and Inside potential of the OECD countries, and conducting a cluster analysis based on the FOI indices, it shows the different development paths these countries are on.
2. By investigating the variables correlated with the FOI indices, and conducting a factor analysis among them, it identifies a Green growth factor of the Future potential.
3. Finally, by comparing the average Green growth factor scores of the different OECD clusters, it identifies the group of countries that have done the best so far at achieving sustainable development.

The rest of the study is structured as follows: section 2 provides a short literature review of the empirical results on the topic of green growth; section 3 introduces the FOI model, and the method and sources of calculating the FOI indices; section 4 presents the results of the calculations and offers a discussion of the results; section 5 concludes the study.

## 2 LITERATURE REVIEW

Although the literature on green growth is very rich, most of the studies focus on either individual countries, or certain sectors/industries of selected countries. In this review I focus on studies conducted on macroeconomic indicators of the OECD countries.

Szita (2014) in an early study calculates a Greening index for the 34 OECD members. Based on the data available for the early 2010s, she finds that the following countries can be considered as green: Austria, Denmark, Norway, Sweden, Switzerland, Japan, and Iceland.

Chakraborty & Mazzanti (2021) test 33 countries that are now all members of the OECD for the 1971-2015 period. They use methodologies that consider cross-sectional dependence and find that there is a significant positive relationship between per capita economic growth and per capita renewable electricity consumption. Higher per capita growth rates seem to go together with higher rates of renewable energy consumption, but the authors could not detect any Granger causality between the two variables (despite trying many specifications).

Gavurova et al. (2021) include all current 38 OECD members in their analysis. They select 15 indicators related to green growth, calculate an average for period 1 (2000-2009), and period 2 (2010-2019) and compare them. The authors find that overall, there is an improvement from period 1 to period 2. A cluster analysis is also conducted, splitting the countries into 6 (period 1), and 7 (period 2) clusters according to the indicators. The countries' cluster placement more or less corresponds to their development level.

Wang et al. (2019) conduct a dynamic panel regression analysis based on the data of the industrial sector of 24 OECD countries for the 2004 and 2010 period to test the impact of environmental regulation policy on green productivity growth. They find an inverted U-shaped relationship, so environmental regulation seems to hurt productivity growth after a certain stringency level. After decomposing the productivity growth, the catching-up effect is found to be the primary source of green productivity growth, which partially explains why some of lesser developed countries (the Czech Republic, Korea, Poland, and the Slovak Republic) of the OECD have the highest green total factor productivity growth. The other group of countries doing well in green productivity growth consists of Finland, Sweden, and France.

Shen et al. (2017) also measure green productivity growth, using a sample of 30 OECD countries over the period of 1970 – 2011. Their results show that green productivity has grown faster than the traditional Total Factor Productivity (TFP) would suggest, because carbon emission drops in periods of downturn. Improvements in technical and structural efficiency contribute to green growth from 1971 – 2000, which is in line with the results of Wang et al. (2019) who also found technical efficiency as the main driver, but for the remainder of the period technological progress seems to have the strongest impact. Shen et al. only provide data for three groups of countries (OECD Americas, OECD Asia-Oceania, and OECD Europe); out of these three groups OECD Europe has the lowest trend in carbon emissions, but this is also the region that grows the slowest over the inspected period.

Huang et al. (2021) take a slightly different approach and focus on the so called 3E trilemma (achieving energy security, economic development, and environmental protection at the same time). The authors set up a complex measurement method based on several indicators and use a sample of 34 OECD countries over the 2000 to 2015 period to test the relationship among the three pillars of the trilemma. Huang et al. could not detect a significant relationship between energy security and decoupling, but they find a group of five countries that have done well both in energy security and in decoupling economic growth from carbon emission. Australia, Switzerland, Germany, Denmark, and Sweden are the five members of this group.

Ates & Derinkuyu (2021) incorporate a number of indicators in their analysis, measuring the economic, social, and environmental aspects of the OECD countries. The authors create a single indicator by synthetizing the different variables using multivariate I-distance approach, and find that Sweden, Luxemburg, Norway, and Denmark are the top performers in green growth.

## 3 DATA AND METHODS

### 3.1 *Data and sources*

My analysis included a sample of 38 OECD countries and incorporated a total of 95 variables measuring the level of socio-economic development. The latest available data was used, which means that most of the values belong either to 2020 or 2019.

The data was obtained from the following sources:
1. OECD.Stat: https://stats.oecd.org/

2. WEF Global Competitiveness Report (Schwab 2019)
3. IMF World Economic Outlook Database, April 2021 Edition: https://www.imf.org/en/Publications/WEO/weo-database/2021/April
4. World Bank Doing Business database: https://www.doingbusiness.org/en/doingbusiness
5. Solability Sustainable Intelligence: https://solability.com/
6. WHO the Global Health Observatory: https://www.who.int/data/gho/data/indicators
7. Global Footprint Network: https://www.footprintnetwork.org/
8. Trading Economics: https://tradingeconomics.com/
9. ETS TOEFL results: https://www.ets.org/

### 3.2 Method: the FOI model

This study uses the FOI model to evaluate the development paths taken by the OECD countries and then identifies a Green growth factor that is correlated with the Future potential, the first of the three pillars of FOI. A detailed description of the method is available in Bartha & Gubik (2014), here only a short summary is provided.

Table 1. Variables used to calculate the F, O, and I indices
Source: own work based on Bartha & Gubik 2014

| Index | Variables |
| --- | --- |
| F-index | 1. Global Sustainable Competitiveness Index (Solability, 2020) |
| | 2. Cooperation in labour-employer relations (WEF-GCI, 2019) |
| | 3. Flexibility of wage determination (WEF-GCI, 2019) |
| | 4. Electricity supply quality (WEF-GCI, 2019) |
| | 5. Total expenditure on educational institutions (OECD, 2017) |
| | 6. Elderly (65 and above) population (OECD, 2020) |
| | 7. Renewable energy (OECD, 2019) |
| | 8. Life expectancy at birth (OECD, 2020) & Healthy life expectancy (WHO, 2019) |
| | 9. Ecological Footprint (GFT, 2020) |
| | 10. R&D expenditures & Patent applications (WEF-GCI, 2019) |
| | 11. 15-year-old students who are not low achievers (OECD, 2018) |
| O-index | 1. Exports+Imports/GDP*2 (OECD, 2020) |
| | 2. Country credit rating (TE, 2020) |
| | 3. Soundness of banks (WEF-GCI, 2019) |
| | 4. Exchange rate stability (IMF, 2019) |
| | 5. TOEFL iBT® Total and Section Score Means (ETS, 2019) |
| I-index | 1. Budget transparency & Burden of government regulation (WEF-GCI, 2019) |
| | 2. Better life index (OECD, 2018) |
| | 3. General government revenue (IMF, 2020) |
| | 4. Assets in pension funds and all retirement vehicles (OECD, 2020) |
| | 5. Gross domestic product per capita & Gross domestic product percent change (IMF, 2020) |
| | 6. Financing of SMEs (WEF-GCI, 2019) |
| | 7. Labour market Flexibility (WEF-GCI, 2019) & Labour force (OECD, 2020) |
| | 8. Ease of finding skilled employees (WEF-GCI, 2019) |

The FOI model is developed on the assumption that three main dimensions determine the development path of economies. The future potential considers the long-term competitiveness of the economy; the outside potential determines the current world market position of the economy; while the inside potential summarizes factors that are crucial for the current well-being of the community. The FOI model assigns several variables to each of the three potentials and following some transformations (all variables are recoded to a 1-7 scale using a minmax method) they can be used to calculate the F, O, and I indices. Table 1 shows all the variables used for the calculation of the three indices.

Once the FOI indices are obtained SPSS is used to derive clusters of OECD countries and factors of the F, O, and I potentials. The hierarchical cluster analysis generates clusters of the 38 countries according to their F, O, and I indices and these clusters can be interpreted as different development strategies or paths. During the factor analysis I select a large number of variables that are correlated with one of the three indices and generate 2 factors for each index. These factors can then be used to provide a more sophisticated description of the OECD clusters. Green growth is one of the factors being connected to the F index.

4 RESULTS AND DISCUSSION

The F, O, and I indices of the OECD countries are included in Table 2. All components were transformed to a 1-7 scale, where 1 is the worst, and 7 is the best value of the indicator. In this paper I focus on the F-index, since it is correlated with many of the variables that are typically used to measure green growth.

Table 2. The F-, O-, and I-index of the OECD countries in 2020
Source: own calculations

| Country | F-index | O-index | I-index |
|---|---|---|---|
| Australia | 3.80 | 5.34 | 4.63 |
| Austria | 4.42 | 5.08 | 3.92 |
| Belgium | 3.82 | 4.87 | 3.59 |
| Canada | 4.00 | 4.93 | 4.64 |
| Chile | 3.65 | 3.90 | 3.78 |
| Colombia | 3.17 | 2.68 | 3.12 |
| Costa Rica | 3.31 | 3.65 | 1.96 |
| Czech Republic | 3.75 | 4.18 | 3.25 |
| Denmark | 4.92 | 5.01 | 4.71 |
| Estonia | 4.16 | 4.73 | 3.61 |
| Finland | 4.63 | 5.07 | 4.95 |
| France | 4.16 | 4.30 | 3.53 |
| Germany | 4.36 | 4.70 | 4.49 |
| Greece | 3.29 | 2.86 | 1.94 |
| Hungary | 3.08 | 4.40 | 2.61 |
| Iceland | 5.34 | 4.23 | 4.96 |
| Ireland | 4.27 | 4.60 | 4.95 |
| Israel | 4.52 | 4.59 | 4.10 |
| Italy | 3.54 | 3.53 | 2.66 |
| Japan | 4.67 | 3.72 | 4.11 |
| Korea | 4.30 | 4.28 | 3.77 |
| Latvia | 3.51 | 4.21 | 3.40 |
| Lithuania | 3.63 | 4.34 | 3.62 |
| Luxembourg | 3.80 | 6.11 | 4.61 |
| Mexico | 3.04 | 4.12 | 3.26 |
| Netherlands | 4.27 | 5.27 | 5.33 |
| New Zealand | 4.54 | 5.08 | 4.78 |
| Norway | 4.70 | 4.87 | 4.86 |
| Poland | 3.69 | 4.00 | 3.12 |
| Portugal | 3.93 | 3.67 | 3.14 |
| Slovak Republic | 3.40 | 4.76 | 2.92 |
| Slovenia | 3.99 | 4.49 | 3.20 |
| Spain | 3.17 | 4.02 | 3.14 |
| Sweden | 4.93 | 4.93 | 4.56 |
| Switzerland | 5.19 | 5.39 | 5.65 |
| Turkey | 3.14 | 3.16 | 3.07 |
| United Kingdom | 3.85 | 5.33 | 4.66 |
| United States | 3.89 | 5.39 | 5.30 |

After the indices were calculated, I checked the bivariate correlation between the index values and the 90+ variables included in my database. Initially, all variables that were correlated to the index on at least 5% significance level were included in a factor analysis using principal components as an extraction method, and varimax for rotation. Some of the variables were correlated to more than one of the indices, so as I proceeded with the iterations, I aimed at including the variables in only one of the factors. The final iteration for the variables correlated to the F-index ended up with two factors (Tab. 3.), with a KMO value of 0.71. The two factors explain 57.8% of the total variance, which is not great, but acceptable. I also calculated a factor score for the Green growth factor using the regression method; the score for Costa Rica and Switzerland is missing, because no data was available for these countries in at least one of the variable categories included in the factor.

Table 3. The two factors of the Future potential (rotated components matrix)
Source: own calculations

| Factor | Variable | Component 1 | Component 2 |
|---|---|---|---|
| Government quality | Efficiency of legal framework in settling disputes | 0.921 | 0.21 |
| | Property rights | 0.902 | 0.29 |
| | Government ensuring policy stability | 0.833 | 0.299 |
| | Strength of auditing and accounting standards | 0.83 | 0.155 |
| | Share of population with tertiary education | 0.787 | 0.022 |
| | Patent applications per million pop. | 0.762 | -0.159 |
| | Incidence of corruption | 0.756 | 0.312 |
| | R&D expenditures | 0.746 | -0.109 |
| | Life expectancy at birth | 0.638 | 0.363 |
| | Contracting with Government | -0.518 | 0.19 |
| | Total expenditure on educational institutions | 0.442 | 0.167 |
| Green growth | Production-based $CO_2$ productivity | 0.02 | 0.713 |
| | Emissions priced above EUR 30 per ton of $CO_2$ | 0.109 | 0.707 |
| | Renewable energy | 0.072 | 0.685 |
| | Population connected to public sewerage | 0.068 | 0.595 |

*Extraction Method: Principal Component Analysis.*
*Rotation Method: Varimax with Kaiser Normalization.*
*a Rotation converged in 3 iterations.*

The cluster analysis was the next step. Countries were sorted according to their F-, O-, and I-index values, hierarchical cluster analysis was applied and between-group linkage was used as the cluster method. I intentionally went for a high number of clusters (11), so that smaller nuances could be detected as well. Table 4 contains the different clusters that were derived, and Figure 1 shows their position along the Future, Outside, and Inside dimensions.

Table 4 also includes the Green growth factor score for the 11 different clusters. Clearly, green growth performance is very heterogeneous even among the most developed countries. Countries in the Welfare-participatory cluster, and some other outliers do relatively well, while the Market-oriented and Statist clusters perform poorly.

The difference between cluster 1 and 2 is particularly striking, as these two clusters (with the exception of three outlier countries: Luxembourg, Iceland, and Switzerland) are typically the top performers in almost all other factors. They have the best scores in both factors of the Inside potential (Human capital and Governance), as well as in the only clearly distinguishable factor of the Outside potential (FDI readiness).

Table 4. Clusters according to their FOI-indices
Source: own calculations

| Nr. | Name | GG fac. score | Members |
|---|---|---|---|
| 1 | Market-oriented | -0.47 | Australia, Canada, Netherlands, United Kingdom, United States |
| 2 | Welfare-participatory | 0.60 | Austria, Denmark, Finland, Germany, Ireland, Israel, New Zealand, Norway, Sweden |
| 3 | Statist 1 (welfare) | -0.52 | Belgium, Estonia, France, Korea, Slovenia |
| 4 | Statist 2 (protectionist) | -0.04 | Chile, Czech Republic, Italy, Latvia, Lithuania, Mexico, Poland, Portugal, Spain |
| 5 | Laggard 1 (rising) | -0.49 | Colombia, Turkey |
| 6 | Laggard 2 (falling) | 0.18 | Costa Rica, Greece |
| 7 | Statist 3 (open) | -0.65 | Hungary, Slovak Republic |
| 8 | Iceland | 2.34 | Iceland |
| 9 | Japan | -1.45 | Japan |
| 10 | Luxembourg | 1.08 | Luxembourg |
| 11 | Switzerland | n/a | Switzerland |

In fact, Cluster 1 (Market-oriented) and Cluster 2 (Welfare-participatory) are very similar to each other in almost every aspects of the FDI model (see Fig. 1.). Their Outside and Inside potentials are similar (with the Market-oriented countries having a slightly higher average in the O-index, and the Welfare-participatory countries a slight edge in the I-index), and they are top per-

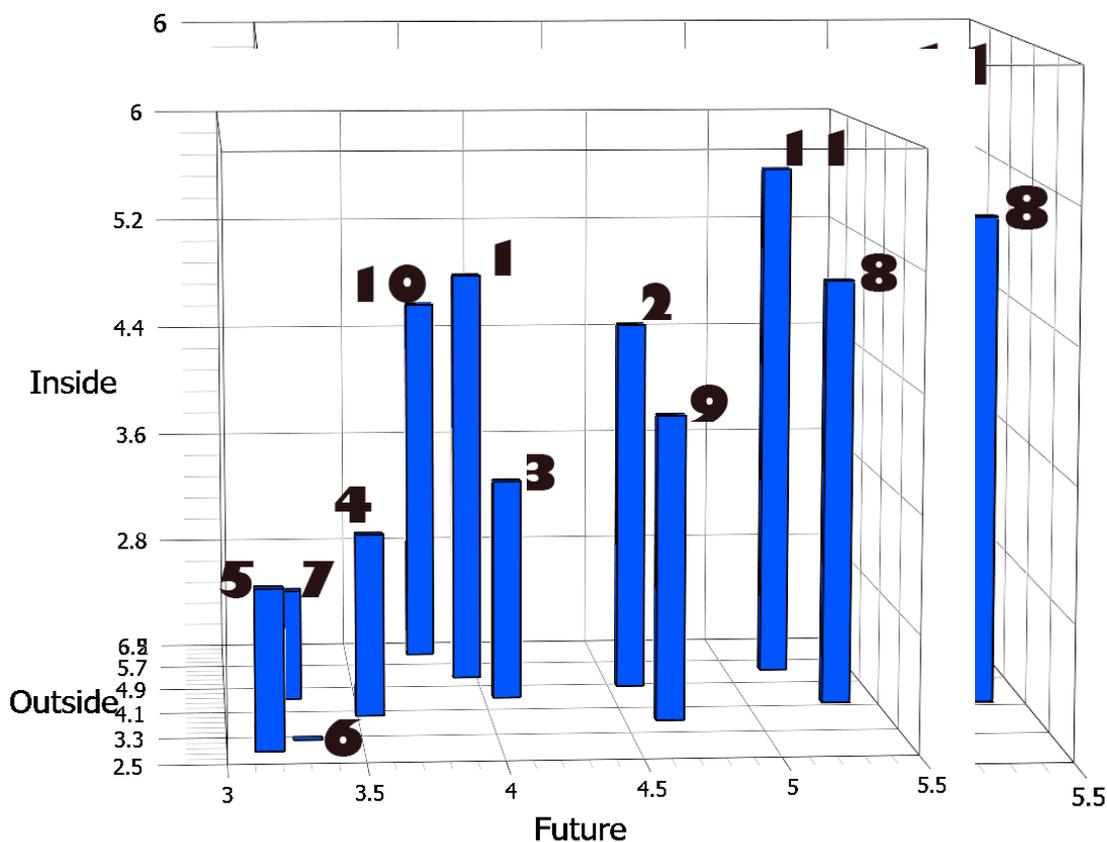

Figure 1. Clusters based the FOI model – an 11-cluster solution
Source: own calculations

When looking for differences, we can check the FDI readiness factor (variables correlated with the O-index), where Cluster 1 has a score of 0.83, while Cluster 2 only scores 0.53, but both of these values are way higher than the factor score of any other multi-country cluster, and they are at the same level as Luxembourg (0.73, very close to Cluster 1's average), and Switzerland (0.54, almost identical to Cluster 2's average).

Based on all the FOI-related calculations, the Market-oriented and the Welfare-participatory countries have the best indicator values within the OECD (again, three, similarly well-performing outliers need to be mentioned here: Iceland, Luxembourg, and Switzerland). They have the highest factor score in Human capital development and Governance (the Market-oriented group having a slight, 0.1-point edge); they also have the highest factor score in FDI readiness (the Market-oriented group leading by about 0.3); and they have the highest F-index as well (although the lead of the Welfare-participatory group is considerable in this area, and the Market-oriented cluster is overtaken by the Statist 1 group as well). So, the only clear difference among the top performers comes from the Green growth factor: the Welfare-participatory countries are on the top of the list (if we disregard Iceland and Luxembourg), while the Market-oriented countries are very close to the bottom of the list, with most of the Statist groups members lining up between them.

## 5 CONCLUSION

In this study I calculated the 2020 FOI indices for the 38 OECD member countries. The cluster analysis conducted with these indices resulted in 11 clusters. Four of these clusters are single-country outlier ones, while the majority of the countries can be found in Cluster 1 (Market-oriented model), Cluster 2 (Welfare-participatory model), and Clusters 3-4 (Statist model). The Market-oriented and the Welfare-participatory clusters are the best performing ones in all aspects measured by the FOI model. The only significant difference between the two can be detected in the Green growth performance, where Cluster 2 countries do really well, while Cluster 1 members score quite low. The countries that I found to do well in green growth are the following:
- members of the Welfare-participatory cluster: Austria, Denmark, Finland, Germany, Ireland, Israel, New Zealand, Norway, Sweden;
- outlier countries: Iceland, and Luxembourg.

According to their Green growth factor score, Norway (2.6), Iceland (2.3), and Luxembourg (1.1) are the top three countries in green growth within the OECD. No factor score was calculated for Switzerland, because at least one data point was missing from the Swiss dataset, but it is likely that they would also be close to the top of this list.

One obvious conclusion that can be drawn from this analysis is that countries with stronger welfare and statist roots have been doing better so far in green growth. This could be the result of their experience in regulating the economy. Wang et al. (2019) has found that stricter environmental regulation tends to go together with better green growth performance, although they have also found evidence for an inverted U-shaped relationship.

Some of the studies cited in the literature review have also named top performer countries. Szita (2014) mentioned Austria, Denmark, Norway, Sweden, Switzerland, Japan, and Iceland; the study of Wang et al. (2019) points to Finland, Sweden, and France; Huang et al. (2021) highlight Australia, Switzerland, Germany, Denmark, and Sweden; finally, Ates & Derinkuyu (2021) find Sweden, Luxemburg, Norway, and Denmark to be the best in green growth. The findings of this study seem to be in line with these (although there are some exceptions, such as Australia, Japan or France).

One of the major limitation of this study is that it works with aggregates of macro statistics, and the aggregation can mask some of the important details. The United Kingdom, for example, has the 7$^{th}$ highest Green factor score in the OECD, but it was still put in the low scoring Cluster 1. Switzerland, on the other hand, would have probably got a high score in green growth, but no factor value was calculated for them because of a single missing data point and the nature of the method selected. Additionally, the Green growth factor is made up of 4 variables, but it is obvious that a wider range of indicators would probably do a better job at evaluating the green growth performance.

One way to improve on these results is the addition of further variables to the analysis and additional research to complete some of the datasets with missing data points (the factor analysis can be distorted by missing data, since one missing point can disqualify the whole country from the analysis). A case study of the well-performing countries could also provide valuable information for policy makers.